\documentclass[manuscript,times]{aastex61}

\newcommand\aastex{AAS\TeX}

\newcommand       \mum           {\,{\rm \mu m}}

\newcommand       \AV           {A_V}

\received{XXX}
\revised{XXX}
\accepted{XXX}
\submitjournal{The Astronomical Journal}

\shorttitle{\aastex\ Relation of Silicate and SiO maser}
\shortauthors{Liu \& Jiang}

\begin{document}

\title{On the relation of silicates and SiO maser in evolved stars}

\correspondingauthor{Biwei Jiang}
\email{bjiang@bnu.edu.cn}

\author{Jiaming Liu}
\affiliation{Department of Astronomy\\
Beijing Normal University \\
Beijing 100875, P. R. China}

\author{Biwei Jiang}
\affiliation{Department of Astronomy\\
Beijing Normal University \\
Beijing 100875, P. R. China}



\begin{abstract}

The SiO molecule is one of the candidates for the seed of silicate dust in the circumstellar envelope of evolved stars, but this opinion is challenged. In this work we investigate the relation of the SiO maser emission power and the silicate dust emission power. With both our own observation by using the PMO/Delingha 13.7-m telescope and archive data, a sample is assembled of 21 SiO v=1,J=2-1 sources and 28 SiO v=1,J=1-0 sources that exhibit silicate emission features in the ISO/SWS spectrum as well. The analysis of their SiO maser and silicate emission power indicates a clear correlation, which is not against the hypothesis that the SiO molecules are the seed nuclei of silicate dust. On the other hand, no correlation is found between SiO maser and silicate crystallinity, which may imply that silicate crystallinity does not correlate with mass loss rate.

\end{abstract}




\section{Introduction} \label{sect:intro}

Silicate is the most important species of oxygen-bearing dust that is believed to be formed in the circumstellar envelope of AGB (Asymptotic Giant Branch) stars, then evolve, be destroyed and reformed in the interstellar space \citep{Liu2014}. Chemically, silicates are made up of cations and silicic acid radical ions ($\rm {SiO_{4}}^{4-}$) or ($\rm {SiO_{3}}^{2-}$). In the circumstellar envelope, the most abundant cations turn out to be of Fe and Mg. Consequently, silicates are divided into olivine (Fe$_{2x}$Mg$_{2(1-x)}$SiO$_{4}$) and pyroxene (Mg$_{(1-x)}$Fe$_{x}$SiO$_{3}$) where $0 \leq x \leq1$. With $x=0$ or $x=1$, the corresponding four chemically distinct types of silicate dust are specifically fayalite(Fe$_{2}$SiO$_{4}$), ferrosilite(FeSiO$_{3}$), forsterite(Mg$_{2}$SiO$_{4}$) and enstatite(MgSiO$_{3}$) \citep{Liu2014}.

Silicate dust is widely detected in astronomical objects via their spectral features. The earlier detections are based on the 9.7 and 18$\mum$ features, which are wide, smooth and featureless and identified as from the Si-O stretching and O-Si-O bending mode respectively of amorphous silicate. The IRAS/LRS detected such spectral features in over 4000 O-rich stars \citep{Kwo2011}. With the increase of spectrometer sensitivity and spectral resolution, the ISO/SWS for the first time detected many narrow spectral features in evolved O-rich stars that are identified as from crystalline silicates \cite[]{Wat1996}. Afterwards, the crystalline silicates are found in various types of objects in the ISO/SWS and Spitzer/IRS spectrum \cite[]{Mol2002a,Mol2002b,Mol2002c,Gie2008}. The narrow features of crystalline silicates appear mainly in seven complexes at approximately 10, 18, 23, 28, 33, 40 and 69$\mum$ \citep{Mol2002b}.

The formation mechanism of amorphous and crystalline silicates is still unclear in spite of numerous detections of the spectral features in the circumstellar envelope of O-rich evolved stars. Dust condensation generally takes two steps \cite[]{Fer2001}. The first step is the formation of tiny seed nuclei probably of the order of a few nanometers, which will serve as centers for later growth to macroscopic grains. The material from which the seed nuclei are formed may either be the same material as the final dust grains (homogeneous dust formation) or be a different material from the final dust grains (heterogeneous dust formation). For carbon grains, there is agreement that the dust grains would experience a heterogeneous growth on the seed nuclei TiC (or ZrC) \cite[]{Ber1996}. Meanwhile, the situation with respect to silicate dust is controversial. It has been long known that homogeneous silicate dust formation is not possible and requires the formation of some different kind of seed particle \cite[]{Don1978,Gai1986,Gai1998a,Gai1998b}. \citet{Gai1998a} have proposed TiO$_{\rm 2}$ as the seed nuclei from theoretical considerations. Corundum(Al$_{2}$O$_{3}$) was also suggested as the seed nuclei of silicate dust formation but met challenge since small aluminium clusters have low bond energy and form only at very low temperatures \citep{Cha1998}. Considering the transparency of the corundum and full Mg-bearing silicate dust in the optical and near-infrared bands \citep{Koi1993,Kem2001}, they were thought to be too transparent to be effectively accelerated away from the star by radiation pressure. But \citet{Nor2012} pointed out that large grains of these two kinds of dust might form in the close vicinity of a few radii to the star via photon scattering rather than absorption and deduced the possibility of corundum-core and silicate-mantle dust. Afterwards, \citet{Hof2016} constructed a Dynamic Atmosphere and Radiation-driven Wind models based on Implicit Numerics (DARWIN) and studied the condensation of Al$_{2}$O$_{3}$ and silicate dust in the vicinity of a few radii from AGB stars. Their work shows that the Al$_{2}$O$_{3}$ dust may condense at a temperature of $\sim 1400\,{\rm K}$ in a vicinity of 2 radii from the star and may serve as the seed nuclei to form a Al$_{2}$O$_{3}$ core and silicate mantle dust(see \citealt{Hof2016}, and reference therein), but under the condition that some impurities due to a few percent of transition metals (e.g., Cr) be incorporated. \citet{Wit2007} also found the connection of SiO maser with corundum by observing S Ori which found that both showed the phase dependence, meanwhile this star show no evidence for hot circumstellar silicate dust. Therefore, corundum may be a strong candidate for the silicate nucleation, but still with some condition and unexplained fact.

Silicon monoxide (SiO) has been discussed several times whether it can serve as the seed nuclei of silicate dust grains. There are reasons supporting this suggestion. Condensation of SiO is the most obvious initial step in silicate formation \cite[]{Nut1982}, since SiO is one of the most abundant of all the gas-phase species of refractory elements in oxygen-rich environment. Besides, a rather stable condensed phase exists with chemical composition. On the other hand, early laboratory experiments showed that the SiO condensation commences at a temperature $\sim$ 600\,K \cite[]{Nut1982,Gai1986}, much lower than the observed condensation temperature of circumstellar silicate that is usually around 900\,K. However \citet{Gai2013} re-measured the vapour pressure of solid SiO which is used to calculate the condensation temperature of SiO. Their new value of the SiO condensation temperature is increased to about 700\,K (depending on the mass loss rate) and much closer to the silicate dust condensation temperature derived from infrared observation. Although the new condensation temperature is still about 100\,K lower than observed, the difference may be explained by the greenhouse effect. This brings about new life of SiO as the seed nuclei of silicate dust formation.

The formation of crystalline silicate dust is even more unclear. The question, which kind of silicate dust form first, amorphous or crystalline silicate, is still under debating. \citet{Gai1999} calculated the condensation sequence of O-bearing molecules in the circumstellar envelopes of O-rich evolved stars, and found that Fe-free olivine will form first due to its high condensation temperature. Their model can solve the question of missing Fe element in the crystalline silicate dust. Due to the efficient absorption in the near infrared,  Fe-containing silicate may only be stable at large radius from the stellar surface \citep{Woi2006,Hof2016}. However, this argument is challenged by lack of laboratory support. Another mechanism needs a conversion from amorphous form through thermal annealing especially in circumstellar disc \citep{Gai1999,Fer2001}. But this model was questioned by \citet{Mol2003} who  argued that the circumstellar disks are after all not an accretion disk, so radial transfer of material may not happen. Other scientists prefer to believe that crystalline silicates are the firstly formed dust. \citet{Tie1998} proposed a model in which the Fe$^{2+}$ ion will penetrate into the crystalline dust and destroy its lattice structure after the condensation of crystalline silicate, which finally leads to formation of amorphous silicate. However, there is no observation of severe reduction of crystalline silicate features.

In this work, we try to study the relation between silicate dust and SiO gas molecules in the evolved stars, which will give some hints on the formation mechanism of silicate dust. SiO molecules are usually traced by their radio lines, either thermal lines from vibrational ground state $v=$0 or maser lines from vibrational excited states $v=$1 or $v=$2. Usually the thermal lines are much weaker than the maser lines \cite[]{Gai2013}. The SiO masers usually occur within a few stellar radii of the stellar surface between the hot molecular inner envelope and the cooler region at 3 $\thicksim$ 5 $R_{\rm star}$ where the dust (especially silicates) forms, and the temperature decreases from 1770K to 1100K \cite[]{Dan1994,Rei1997}.

\section{Sample selection and SiO maser observation} \label{sect:samp}
\subsection{Sample and observation}\label{sect:sa&obs}
The correlation between the intensity of SiO maser and silicate emission would reflect whether SiO molecules are the seed of silicate dust. In order to study the intensity relation, we try to detect SiO maser emission from evolved O-rich stars. The sample of observation is consisted of evolved stars which show emission lines of crystalline silicate in order to investigate whether SiO molecules are related to silicate crystallization. We select the stars from three relatively systematic search for crystalline silicate by \citet{Gie2008}, \citet{Olo2009} and \citet{Jon2012}. For the convenience of calculating the intensity of crystalline silicate emission, we require that the fully processed ISO/SWS spectrum is available. The ISO/SWS spectrum \cite[]{de1996,Kes1996} covers a wider range (2.4-45.4 $\mum$) of wavelength in comparison with the Spitzer/IRS spectrograph (5.2-38.0$\mum$) \citep{Wer2004}  so that the measurement of the intensity of crystalline silicate would be more complete. In addition, the location of the 13.7-m telescope of the Purple Mountain Observatory (PMO) in Delingha, China (PMO/Delingha hereafter) to perform the observation constrains the sources in the northern sky. With the above considerations, a sample of 31 evolved stars is selected, whose position and type are listed in Table 1. It can be seen that the sample is consisted of AGB stars, OH/IR stars, post-AGB stars or PPN, PN, and a few supergiant stars. From the available IRAS measurement at 12, 25 and 60$\mum$, all except two have [12]-[25] ($\equiv2.5*\lg F_{\rm 25}/F_{\rm 12}$ in accordance with \citealt{van1988} ) bigger than 1.5 or [25]-[60] $>$ 0. According to the infrared spectrum either from \emph{ISO}/SWS or \emph{Spitzer}/IRS, the silicate feature(s) around 10$\mum$ is mostly in emission while in absorption for seven objects. In the following it will be shown that the SiO maser is not detected in the sources that exhibit 10$\mum$ absorption except in OH 26.5+0.6. OH 26.5+0.6 is removed for later analysis of the relation between SiO maser and silicate dust because it is optically thick around 10$\mum$.

\begin{figure}
\centering
   \includegraphics[width=7.0cm, angle=0]{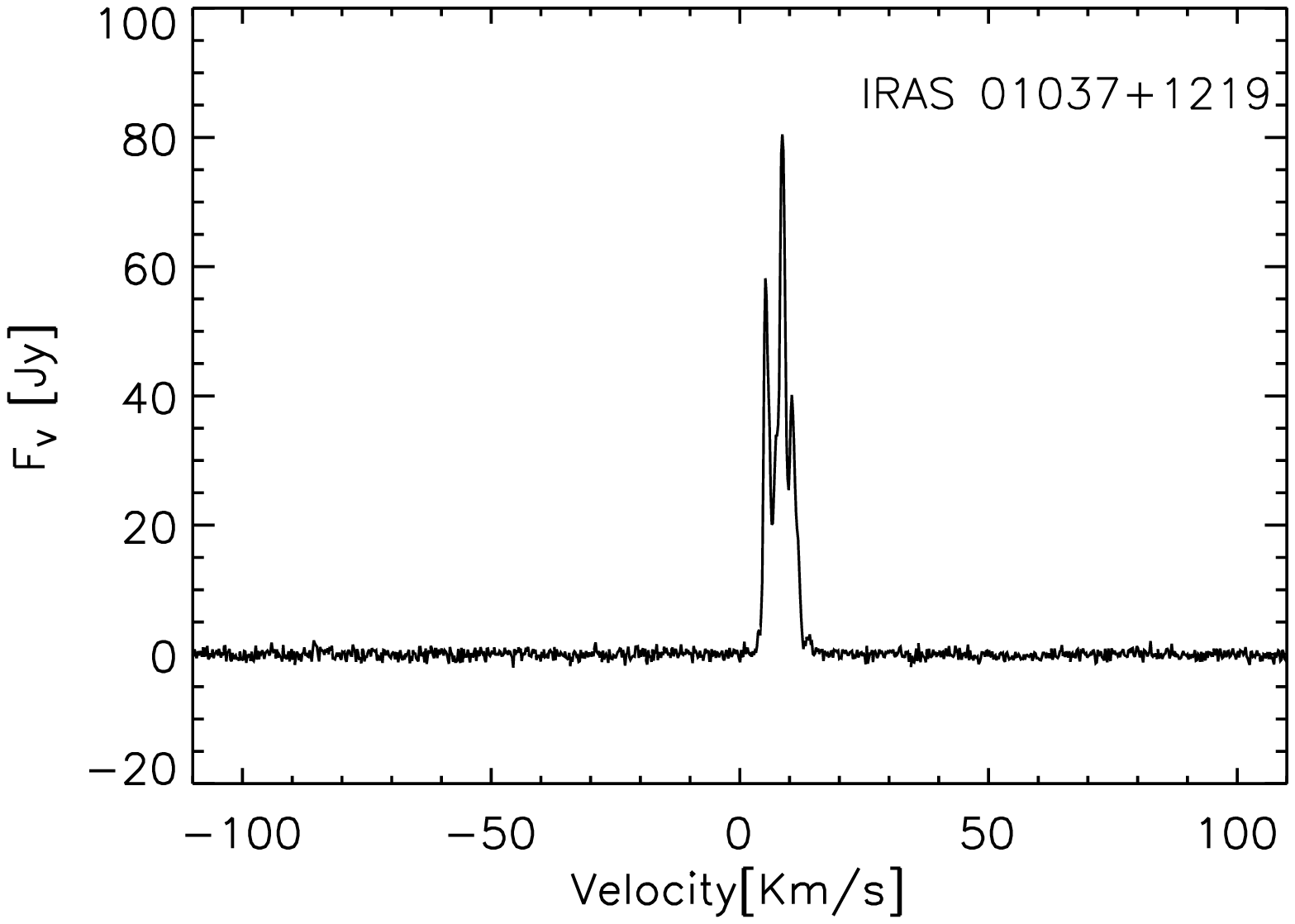}
   \includegraphics[width=7.0cm, angle=0]{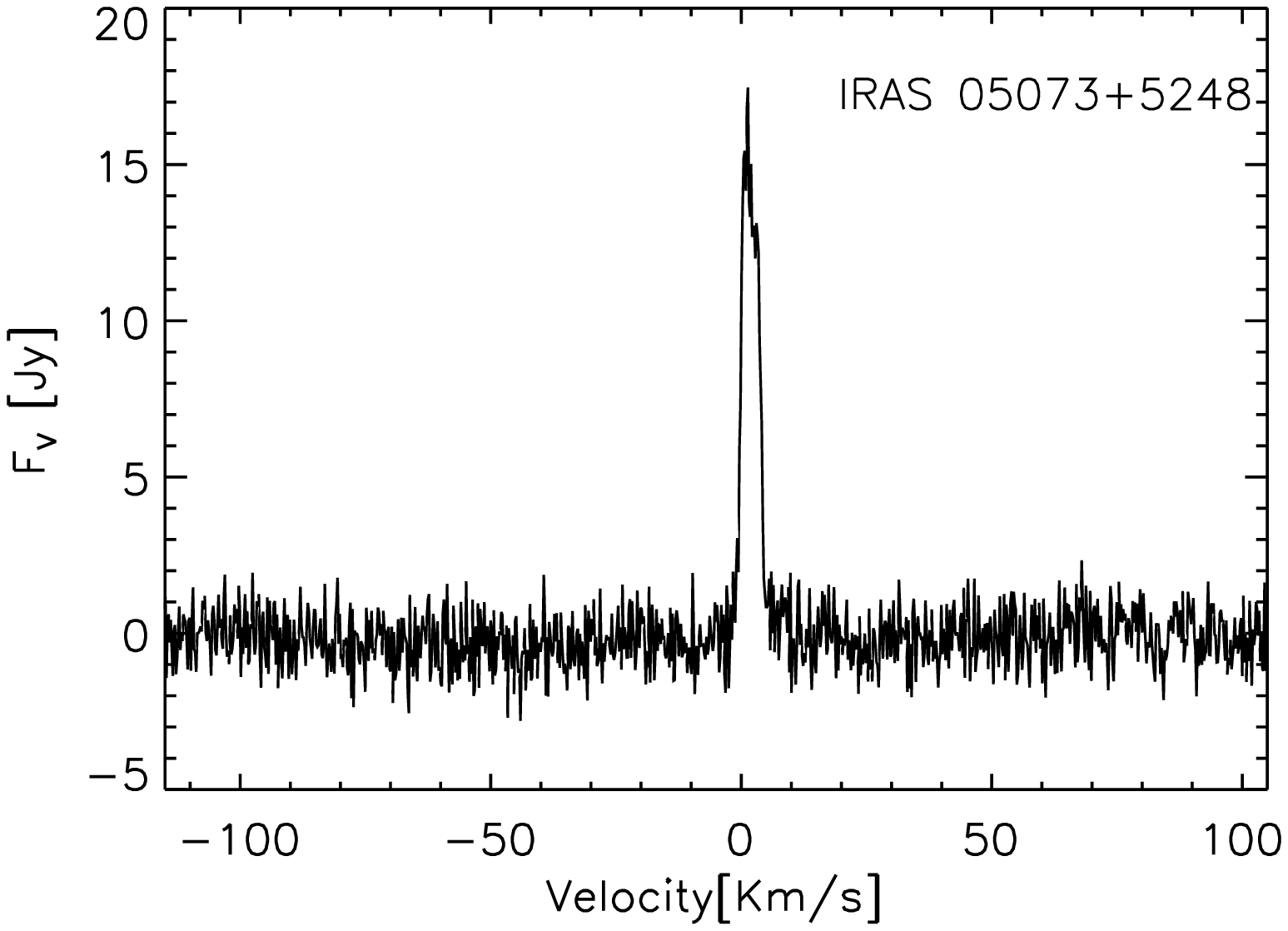}
   \includegraphics[width=7.0cm, angle=0]{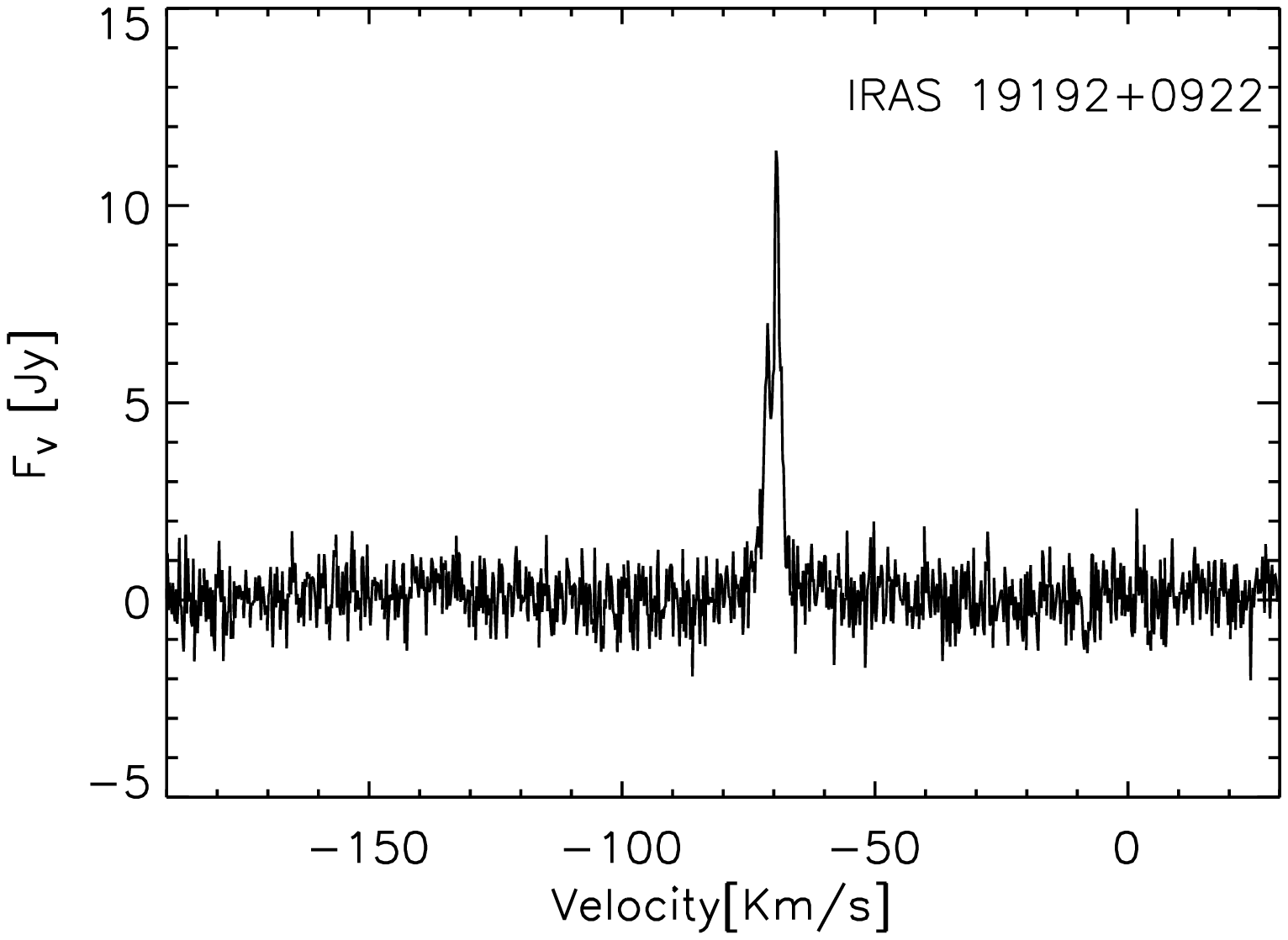}
   \includegraphics[width=7.0cm, angle=0]{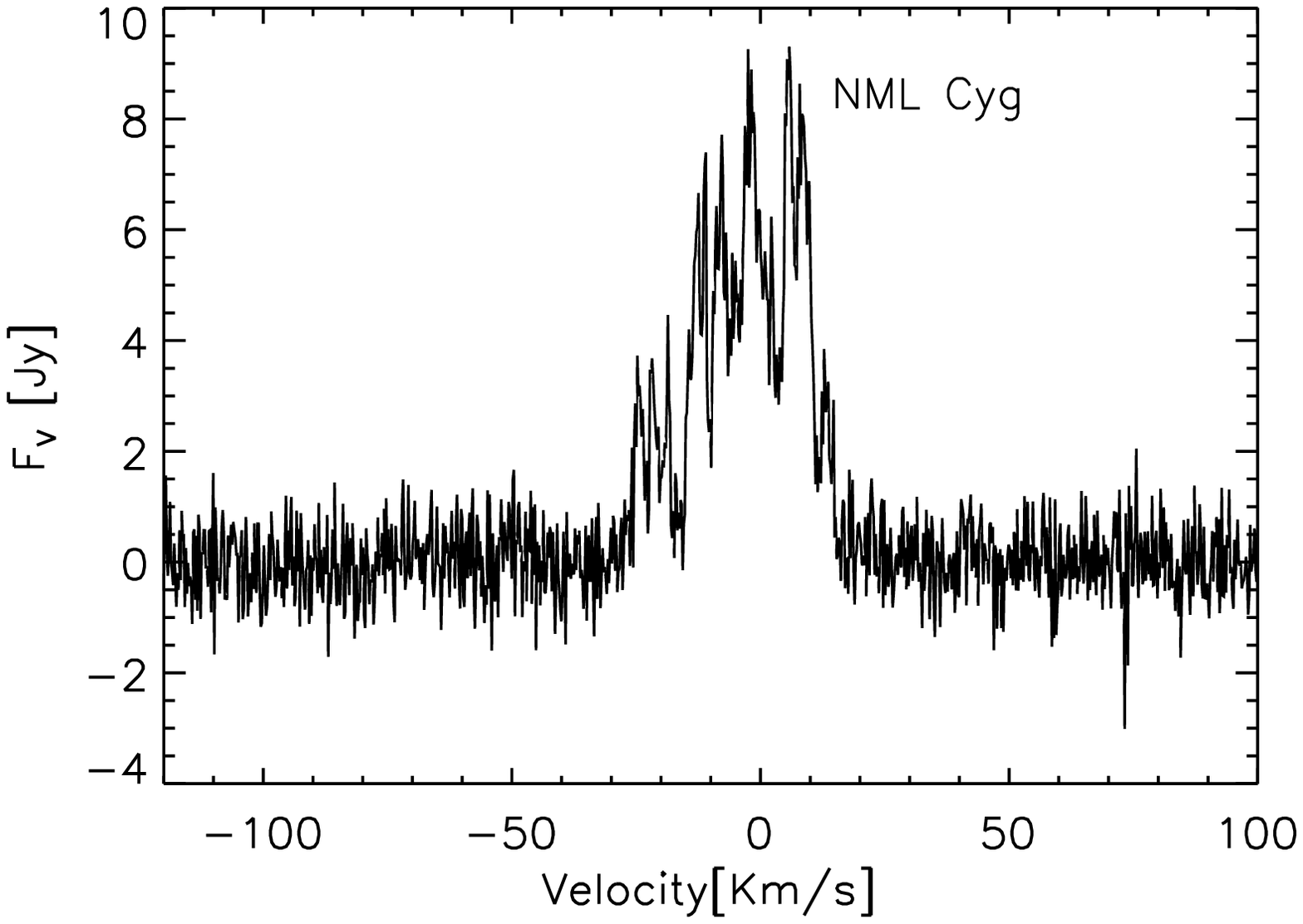}
   \includegraphics[width=8.0cm, angle=0]{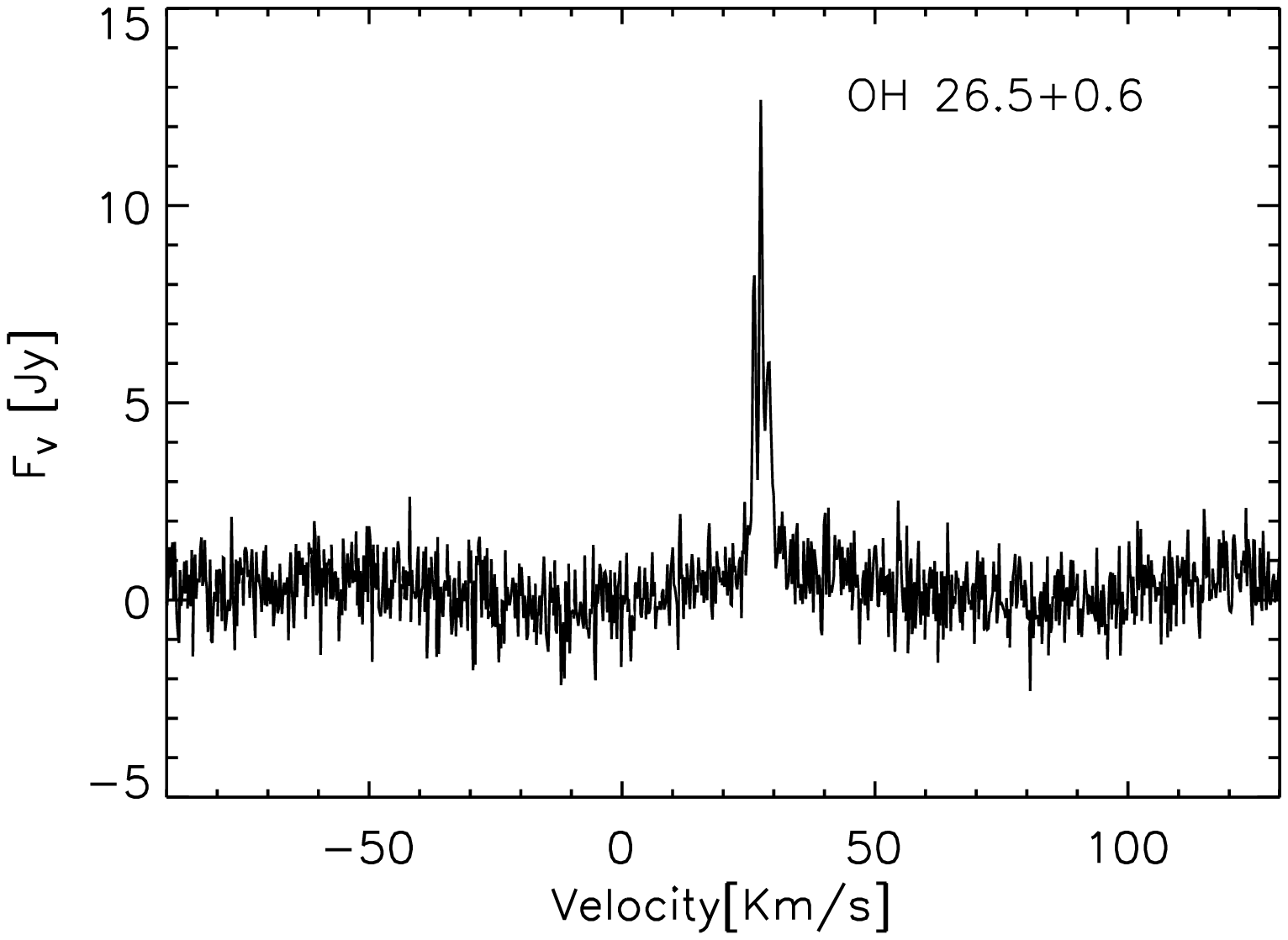}
\caption{The observation result of SiO $v=1,J=2-1$ 86GHz maser emission by the PMO/Delingha 13.7m telescope.}
\label{Fig1}
\end{figure}

The observation was carried out from 2 May 2013 to 2 June 2013 by using the PMO/Delingha 13.7m telescope. The central frequency is tuned to the rest frequency of the SiO $v=1, J=2-1$ line, i.e. 86.243GHz. The observation mode is single-point, with a band width of 1GHz and a frequency resolution of 61kHz corresponding to a velocity resolution of 0.21 km/s. The observations were executed with the low side-band of Beam 2(B2LS) at 86.1 GHz, of which the main beam width was about 61$''$, and the main beam efficiency ($\eta_{\rm {mb}}$) was 0.557 for the $^{28}$SiO $v=1, J=2-1$ line. The typical rms noise level of B2LS is usually better than 0.09\,K at the 4$\sigma$ level with an on-source exposure time of 1800s as the system temperature was normally around 160\,K  (see Status Report on the 13.7 m MM-Wave Telescope for the 2013-2014 Observing Season). All the raw data were processed with the GILDAS package \citep{Gil2013}. It should be noticed that any results presented in the Figures and Tables have been converted to the intensity unit Jansky (Jy) with a conversion factor calculated by the formula: $\frac{2k}{A\cdot \eta}$, where $k$ is the boltzmann constant, $A$ is the physical area of the antenna and $\eta$ is the aperture efficiency. Since the aperture efficiency during our observation time was 0.461, the scale factor from antenna temperature to Jy turned out to be 40.63. The detailed information of those stars including the equatorial coordinates (J2000), type and the observational results are listed in Table \ref{delingha}. We detected the SiO maser emission from  5 of the 31 sample stars. They are IRAS 01037+1219 (Oxygen rich AGB star), IRAS 05073+5248 (OH/IR star), IRAS 19192+0922(Proto-Planetary Nebula), OH 26.5+0.6 (OH/IR star, removed for later analysis due to its absorption around 10$\mum$) and NML Cyg (Red Supergiant) (see Figure \ref{Fig1}). These stars were previously detected and our observation confirmed that they are SiO maser emitters. Due to the relatively small aperture of the 13.7-m antenna, no new detection was found. On the other hand, the SiO maser is variable and its intensity is then phase dependent (e.g. \citealt{Gra2009}), thus only a fraction of the sample can be detected at a given epoch. Although it would be better to analyze the relation of SiO maser detection with phase, accurate periods are not known for many of the sample stars, and in some cases they may be irregular.

\subsection{Sources from literatures}\label{sect:sa&ext}

As the SiO maser detection rate is lower than expected, our sample needs to be expanded in order to obtain a reliable statistical result. We assembled the SiO maser sources from published papers. From numerous detections of SiO maser, the objects are selected by the following criteria: (1) with the ISO (Infrared space Observatory) spectral data, which have already been processed by Sloan et al. (2003) in a uniform manner, to guarantee the quality of the infrared spectrum; (2) exhibition of distinguished 9.7 and 18 $\mum$ emission features which indicate silicate dust, to ensure the detection of silicates;  and (3) with clearly SiO maser emission at $v=1,J=2-1$ or $v=1,J=1-0$. Consequently, the sample has 21 sources with the  SiO $v=1,J=2-1$ line and 28 sources with the SiO $v=1,J=1-0$ line emission. Finally, all the intensity units of the SiO maser emission are converted into Jy for convenience of comparison (see Table \ref{tbl2}). For the sources from the work of \citet{Cho2009}, \citet{Kim2010} and \citet{Cho2012}, the conversion factor is 54.0, 2.5175 and 13.29 respectively. In combination with the four sources from our observation, the sample all show emission features at the wavelengths of silicate implying optically thin case.


\section{Calculation of silicate dust emission} \label{sect:met}

The emission power of silicate dust is calculated by the PAHFIT package \citep{Smi2007} incorporated into the IDL software. Following the method of \citet{Liu2017}, we employed the CDE-profile of \citet{Dor1995} to characterize the emission feature of amorphous silicates. Though the PAHFIT code was originally designed on the purpose of simulating the features of PAH molecules, the code was lately modified and can then be used to fit the features of silicate dust. The advantage of this code is that one can figure out the features' information precisely and avoid inducing uncertainties brought in by dust temperatures for calculation of silicate crystallinity \citep{Liu2017}. To make sure that we have recognized the crystalline silicate's features correctly, we also consulted the work of \citet{Mol2002b}, and only considered the crystalline silicate features that have been identified. An example is showed in Figure 2. From Figure 2 we can see that this fitting process decompose the ISO/SWS spectrum into (1) a stellar continuum (bright blue line), (2) a dust continuum (orange dashed-dotted lines): $F_{\rm dust\ cont.}$, (3) the amorphous silicate features (red solid line, the sum of the two dotted lines mainly from the 9.7$\mum$ and 18$\mum$ respectively): $F_{\rm am.\ feat.}$, and (4) a series of individual narrow features (blue solid lines). Among the individual features, we only choose those features that have been proved to be from crystalline silicate by \citet{Mol2002b}, took the sum of their fluxes as the emitting fluxes of crystalline silicate dust: $F_{\rm cry.}$. Considering that most of the dust continuum will be ascribed to the amorphous silicate dust \citep{Mol2001, Mol2002a, Mol2002b, Mol2002c, Jia2013}, we can consequently attain the emission flux of amorphous silicate dust $F_{\rm am.}=F_{\rm dust \ cont.}+F_{\rm am.\ feat.}$.

Then, crystallinity of silicate, defined as the mass fraction of crystalline silicate dust to total silicate dust, can be approximated by the flux ratio of crystalline silicate dust to silicate dust, i.e. $\eta=F_{\rm cry.}/(F_{\rm cry.}+F_{\rm am.})$ \citep{Liu2017}. Though the flux ratio $\eta$ can not serve the same as the mass ratio of crystalline to amorphous silicate dust, using this ratio as the indicator of crystallinity of silicates is reasonable. The emission flux $F_{\nu} \propto m_{\rm dust}{\rm \kappa_{abs}} B_{\nu}(T_{\rm dust})$, which means that the emission flux ratio would represent the mass ratio if the crystalline and amorphous silicate dust share the same emissivity. As \citet{Kem2001} pointed out, the emissivity of crystalline silicate is comparable to amorphous silicate in mid-infrared, and weaker in visible and near-infrared. Thus in the optically thin case, where the crystalline silicate dust may be colder than amorphous, the flux ratio $F_{\rm cry.}/(F_{\rm am.}+F_{\rm cry.})$ would be lower than the mass ratio $m_{\rm cry.}/(m_{\rm am.}+m_{\rm cry.})$ and could be regarded as the lower limit of crystallinity. While in the case of high mass loss rate, when the infrared radiation mainly comes from the outer cold dust envelope in the mid-infrared where amorphous and crystalline silicates share similar emissivity and dust temperatures, the flux ratio would then approximate very closely to the dust mass ratio. Notice that most of the sources are within 1 kpc even the furthest one AH Sco is 2.7 kpc from the Earth(see Table \ref{tbl2}), the influence of interstellar dust on the stellar spectra should not be serious. According to recent determination of interstellar extinction in the infrared, the extinction around the 10$\mum$ silicate feature is only about 5 percent of the visual band (see e.g. \citealt{Xue2016}), which converts to less than 5 percent decrease of flux if an average $\AV = 1.0$ mag per kpc is adopted. Moreover, there is no detection of crystalline silicate in the diffuse ISM which set an upper limit of interstellar crystallinity to less than 5\% \citep{Li2001,Li2007} or 2.2\% \citep{Kem2005}. Since the interstellar silicate absorption is very small (amorphous) or little (crystalline), the contamination of interstellar silicate is not taken into account.

For the SiO maser power, we simply integrated its flux intensity over the velocity space and took it as the emission power. The results are listed in Table \ref{tbl2} and Table \ref{tbl3}.

\section{result and discussion} \label{sect:result}

Following the method of Section \ref{sect:met}, the total emission power of the silicate dust and the flux ratio are computed and listed in Table \ref{tbl3}. As discussed above, the flux ratio of crystalline to total silicate could be equivalent to or be the lower limit of the silicate crystallinity. The flux ratio in Table \ref{tbl3} ranges from about 6\% to 29\% with a concentration around 10\%. As \citet{Hen2010} pointed out, The relative abundance of crystalline silicates is in general quite modest, 10\% $-$ 15\%, in the evolved stars. The work of \citet{Liu2017} yielded a similar result, most of their sources showed a crystallinity of $8\%\sim 16\% $ while a range of $5\%-28\%$. Our consistency with these works indicates that the flux ratio indeed represents the mass ratio in these evolved stars.

With the total silicate dust emission power (Table \ref{tbl3} column 2+column 3) and the SiO maser emission power from Table \ref{tbl2} we can easily figure out the correlation between them. However one thing that we should bear in mind is that the emission power of both silicate dust and SiO maser is inversely proportional to the square of distance, which means there should always be correlation between the apparent power of silicate dust and SiO maser power due to this geometrical effect. In order to eliminate the distance effect and to study the relation between the intrinsic power of silicate dust and SiO maser, both measured powers are normalized to their intensity in the IRAS 60$\mum$ band (column 5 of Table \ref{tbl3}). NML Cyg and U Cep are removed from following analysis due to the lacking of IRAS 60$\mum$ measurement.

Figure \ref{Fig3} shows that the emission power of silicate dust is clearly correlated with the emission power of SiO maser features with respectively the Pearson correlation coefficient of 0.60 for SiO $v=1,J=1-0$ and 0.44 for $v=1,J=2-1$. Because the emission power is tightly correlated to the mass for both SiO masers and silicate dust, the correlation of emission power indicates that the abundance of silicate dust is correlated with that of SiO molecules. This relation may imply that SiO is the seed nuclei of silicate dust grains as suggested by \citet{Nut1982} and \citet{Gai1986}. On the other hand, there are environmental factors in addition to the emitter abundance that influence the emission power. For SiO maser, the emission is an amplified maser radiation whose intensity relies on the abundance in a non-linear way. Depending on the pumping mechanism, the maser intensity changes with the velocity gradient for a radiative pumping \citep{Deg1976} and with the density for a collisional pumping \citep{Eli1980}. \citet{Gra2009} modelled the SiO maser in combination with the dynamic atmosphere which found that the infrared radiation of dust plays some role in the maser emission. This result implies a correlation between the SiO maser emission and dust radiation. Since the silicate dust is the dominant component of dust in the oxygen-rich circumstellar envelope, the present correlation of silicate dust emission and SiO maser power is expected.  To make things more complex, silicate features show both phase-dependent and evolutionary variability \citep{Mon1998,Mon1999}, however, the silicate features vary very mildly and should not bring significant difference. The SiO maser varies as well and in a more irregular way.  Because the method is statistical and the sample spans various phases of variation, variability may not be a serious problem.
SiO masers from the v=1 state are usually found within about 2-4 stellar radii (e.g. \citealt{Rei1997}) and most of the dust characterised as silicate, around stars with SiO masers, is formed just outside this at 5-10 stellar radii (e.g. \citealt{Dan1994}). Thus, gas-phase SiO and silicate dust are found in close proximity.  Therefore, the correlation between SiO maser power and silicate dust emission power may indicate a true connection between SiO molecules and silicate nucleation, but the connection is not very direct since other factors play roles in determining the emission power.

Different from the total emission power of silicate dust, the crystallinity (here defined as $\eta=F_{\rm cry.}/(F_{\rm cry.}+F_{\rm am.})$ ) shows hardly any correlation with the SiO maser emission power (see Figure \ref{Fig4}). With a very small ($<0.1$) Pearson correlation coefficient, the crystallinity neither shows any correlation with the total silicate emission power. What determines crystallinity of evolved stars is an open question, a popularly studied factor is mass loss rate while the conclusion is still controversial. \citet{Jon2012}  examined whether the strength of some specific strong features around 23, 28 and 33$\mum$ correlated with mass loss rate, and did not find any clear quantitative correlation, except a general tendency that stars with high mass loss rate would have higher probability of crystallization. Some people think that the crystalline silicates can only be formed in high mass loss rate stars which have high dust column density \citep{Tie1998, Gai1999}. A latest work of \citet{Liu2017} considered 28 oxygen-rich evolved stars. With PAHFIT code they fitted the ISO/SWS spectra of these sources and deduced the crystalline silicates flux ratios. Together with the dust mass loss rate obtained from SED modelling by 2DUST code \citep{Uet2003}, they investigated the relation between the silicate dust crystallinities and the stellar dust mass loss rates, their result also showed that the silicate's crystallinity is barely correlated with the stellar dust mass loss rate. The above analysis indicates that the SiO maser emission power is correlated with the silicate dust emission that is proportional to the dust mass for the investigated optical thin cases, thus the SiO maser power could be correlated with the dust mass  although the dust temperature would also play important role in determining the dust emission power. Furthermore, the dust mass can be a function of the mass loss rate.  From the fact that there is no correlation between crystallinity and SiO maser, it may be drawn indirectly that crystallinity is not correlated to mass loss rate since it does not correlate with silicate dust mass in the optically thin case.

\section{summary}

In this work we investigated the relation between SiO maser power and silicate dust emission power in order to study whether SiO molecules are the seed nuclei of silicate dust. The observation to search for SiO $v=1, J=2-1$ maser line was performed by using the PMO/Delingha 13.7-m radio telescope and five of 31 sources were detected successfully. In combination with previous observational results, we composed a sample of 21 oxygen-rich evolved stars with SiO $v=1, J=2-1$ line and 28 stars with SiO $v=1, J=1-0$ line. The silicate emission power is calculated by using the PAHFIT code. We found that there is clear correlation between the SiO maser and the silicate's emission power, which is not against the hypothesis that the SiO molecules may serve as the seed nuclei of silicate dust. On the other hand, no correlation is found between SiO maser and silicate crystallinity. 




\acknowledgments
We thank Prof. Aigen Li for helpful discussion and anonymous referee for constructive suggestions. This work is supported by NSFC through Projects 11373015, 11533002, and 973 Program 2014CB845702.
We acknowledge the use of PMO/Delingha 13.7m telescope and the sincere help of the staff.

\startlongtable
\begin{deluxetable}{l|c|c|c|c|c|c|c|c|c}
\tablecaption{List of sources for search of the SiO $v=1,J=2-1$  86GHz maser \label{delingha}}
\tablehead{
\colhead{Name} & \colhead{R.A.} & \colhead{Decl.} & \colhead{Type}&\colhead{A/E$^{1}$} & \colhead{Disc} & \colhead{Det$^{2}$} & \colhead{Sigma} & \colhead{$V_{\rm {LSR}}$} & \colhead{FWHM}\\
\colhead{}&\colhead{}&\colhead{}& \colhead{}&\colhead{}&\colhead{}&\colhead{}&\colhead{K}& \colhead{km/s} & \colhead{km/s}
}
\startdata
IRAS 01037+1219	& 01 06 25.98  & +12 35 53.05	&   O-AGB		    &   E	 & -{-} &  Y & 0.017  & 7.947     & 5.692    \\
IRAS 01304+6211	& 01 33 51.21  & +62 26 53.20	&   OH/IR star      &   E    & -{-} &    & 0.022  &&    \\
TW Cam		    & 04 20 47.62  & +57 26 28.47	&   Post-AGB        &   -{-} & disc$^{a,b}$ &    & 0.016  &&   \\
IRAS 05073+5248	& 05 11 19.44  & +52 52 33.20	&   OH/IR star     	&   E	 & -{-} &  Y & 0.022  & 1.531     & 3.731   \\
IRAS 06034+1354	& 06 06 14.91  & +13 54 19.10	&   Post-AGB		&   -{-} & disc$^{a,b}$ &    & 0.022  &&   \\
IRAS 06072+0953	& 06 09 57.99  & +09 52 31.82	&   Post-AGB        &   -{-} & disc$^{a,b}$ &    & 0.023  &&   \\
SU Gem		    & 06 14 00.02  & +27 42 12.17	&   Post-AGB        &   -{-} & disc$^{a,b}$ &    & 0.021  &&    \\
UY CMa		    & 06 18 16.37  & -17 02 34.72	&   Post-AGB        &   -{-} & disc$^{b}$ &    & 0.024  &&   \\
HD 44179	    & 06 19 58.22  & -10 38 14.71   &   post-AGB       	&   E    & disc$^{b}$ &    & 0.024  &&   \\
HD 45677	    & 06 28 17.42  & -13 03 11.14   &   B[e] star		&   E	 & -{-} &    & 0.023  &&   \\
IRAS 06338+5333	& 06 37 52.43  & +53 31 01.96	&   post-AGB        &   -{-} & disc$^{b}$ &    & 0.021  &&   \\
HD 52961	    & 07 03 39.63  & +10 46 13.06	&   Post-AGB		&   -{-} & disc$^{b}$ &    & 0.022  && \\
HD 161796	    & 17 44 55.47  & +50 02 39.48   &   Post-AGB      	&   -{-} & -{-} &    & 0.023  &&  \\
NGC 6543	    & 17 58 33.42  & +66 37 59.52   &   PN		        &   -{-} & -{-} &    & 0.023  &&  \\
IRAS 18123+0511	& 18 14 49.39  & +05 12 55.70	&   Post-AGB		&   E	 & disc$^{b}$ &    & 0.021  &&  \\
MWC 922		    & 18 21 16.06  & -13 01 25.69	&   Peculiar object	&   A    & -{-} &    & 0.035  &&  \\
MWC 300		    & 18 29 25.69  & -06 04 37.29   &   supergiant		&   A	 & -{-} &    & 0.027  &&  \\
AC Her		    & 18 30 16.24  & +21 52 00.61   &   post-AGB star	&   E    & disc$^{a,b}$ &    & 0.102  &&  \\
OH 26.5+0.6	    & 18 37 32.51  & -05 23 59.20	&   OH/IR star		&   A	 & -{-} & Y  & 0.018  & 27.326     & 4.003  \\
IRAS 18354-0638	& 18 38 06.00  & -06 35 35.00	&   post-AGB		&   -{-} & -{-} &    & 0.044  &&  \\
EP Lyr		    & 19 18 19.55  & +27 51 03.19	&   post-AGB        &   -{-} & disc$^{a,b}$ &    & 0.028  &&  \\
IRAS 19192+0922	& 19 21 36.52  & +09 27 56.50	&   PPN			    &   E 	 & -{-} & Y  & 0.016  & -70.093    & 3.498  \\
IRC+10420	    & 19 26 48.10  & +11 21 16.74   &   post-RSG		&   E	 & -{-} &    & 0.018  &&  \\
IRAS 19283+1944	& 19 30 29.48  & +19 50 41.00	&   PN		     	&   A	 & -{-} &    & 0.020  &&  \\
IRAS 20043+2653	& 20 06 22.74  & +27 02 10.60	&   OH/IR Thick		&   A	 & -{-} &    & 0.017  &&  \\
IRAS 20056+1834	& 20 07 54.62  & +18 42 54.50	&   post-AGB        &   -{-} & disc$^{b}$ &    & 0.012  &&  \\
NML Cyg		    & 20 46 25.54  & +40 06 59.40	&   Red Supergiant	&   A	 & -{-} & Y  & 0.016	& 1.581      & 26.772  \\
IRAS 21554+6204	& 21 56 58.18  & +62 18 43.60	&   OH/IR Thick		&   A	 & -{-} &    & 0.016  &&  \\
IRAS 22177+5936	& 22 19 27.48  & +59 51 21.70	&   OH/IR star    	&   A	 & -{-} &    & 0.018  &&  \\
HD 213985	    & 22 35 27.52  & -17 15 26.89	&   post-AGB        &   E    & disc$^{b}$ &    & 0.017  && \\
IRAS 23239+5754	& 23 26 14.82  & +58 10 54.60   &   PN		        &   E    & -{-} &    & 0.015  && \\
\enddata
\tablenotetext{1}{ ``A'' means absorption in 9.7$\mum$ feature, while ``E'' means emission.}
\tablenotetext{2}{ The observational result of Delingha.}
\tablenotetext{a,b}{ \hspace{0.3cm}Reference: (a) \citet{Gez2015}; (b) \citet{Ruy2006}}
\end{deluxetable}

\startlongtable
\begin{deluxetable}{l|c|c|c|c|c}
\tiny
\tablecaption{The SiO maser sources\label{tbl2}}
\tablehead{
\colhead{Name} & \colhead{R.A.} & \colhead{Decl.} & \colhead{Dist$^{\alpha}$} & \colhead{P$_{J=1-0}$$^{\beta}$} & \colhead{P$_{J=2-1}$$^{\beta}$}\\
\colhead{}&\colhead{}&\colhead{}& \colhead{pc} & \colhead{Jy km/s}&\colhead{Jy km/s}
}
\startdata
AH Sco          & 17 11 17.02  & -32 19 30.71  & 2600$^{1}$ & 112.03$^{a}$      &   -{-}        \\
BU And          & 23 23 39.90  & +39 43 36.92  & 554$^{2} $ & 30.0086$^{a}$     &   -{-}        \\
EP Aqr          & 21 46 31.85  & -02 12 45.93  & 173$^{2} $ & 0.3524$^{a}$      &   -{-}        \\
FP Aqr          & 20 46 36.50  & -00 54 11.10  & 140$^{2} $ & 4.89$^{b}$	    &   -{-}        \\
GY Aql          & 19 50 06.33  & -07 36 52.49  & 1869$^{2}$ & 20.11$^{b}$       &   130.14$^{d}$\\
IRAS 01037+1219 & 01 06 25.98  & +12 35 53.05  & 540$^{3} $ & -{-}              &   302.35$^{e}$\\
IRAS 05073+5248 & 05 11 19.44  & +52 52 33.20  & 510$^{1} $ & -{-}              &   61.38$^{e}$ \\
IRAS 19192+0922 & 19 21 36.52  & +09 27 56.50  & 950$^{3} $ & -{-}              &   36.02$^{e}$ \\
O Cet           & 02 19 20.79  & -02 58 39.50  & 1325$^{3}$ & -{-}              &   447.44$^{f}$\\
R And           & 00 24 01.95  & +38 34 37.35  & 120$^{2} $ & 2.2657$^{a}$      &   -{-}        \\
R Aql           & 19 06 22.25  & +08 13 48.01  & 1273$^{2}$ & 73.00$^{c}$       &   334.90$^{g}$\\
R Aqr           & 23 43 49.46  & -15 17 04.14  & 310$^{1} $ & 225.79$^{b}$      &   236.78$^{h}$\\
R Cas           & 23 58 24.87  & +51 23 19.70  & 181$^{1} $ & 251.25$^{a}$      &   335.42$^{g}$\\
R Hya           & 13 29 42.78  & -23 16 52.77  & 359$^{1} $ & 189.00$^{c}$      &   543.78$^{d}$\\
R Peg           & 23 06 39.17  & +10 32 36.09  &  83$^{2} $ & 29.93$^{a}$       &   -{-}        \\
RR Aql          & 19 57 36.06  & -01 53 11.33  & 480$^{1} $ & 92.29$^{a}$       &   -{-}        \\
RR Per          & 02 28 29.40  & +51 16 17.32  & 758$^{1} $ & -{-}              &   57.24$^{d}$ \\
RT Vir          & 13 02 37.98  & +05 11 08.38  & 816$^{3} $ & -{-}              &   127.98$^{d}$\\
S Per           & 02 22 51.71  & +58 35 11.45  & 230$^{2} $ & 63.72$^{a}$       &   125.21$^{g}$\\
S Scl           & 00 15 22.27  & -32 02 42.99  & 2300$^{4}$ & 4.33 $^{a}$       &   -{-}        \\
S Vir           & 13 33 00.11  & -07 11 41.02  & 362$^{2} $ & 16.44$^{a}$       &   -{-}        \\
SS Peg          & 22 33 58.33  & +24 33 53.98  & 477$^{1} $ & 27.79$^{b}$       &   -{-}        \\
T Cas           & 00 23 14.27  & +55 47 33.21  & 568$^{2} $ & 8.23$^{b}$	    &   -{-}        \\
T Cep           & 21 09 31.78  & +68 29 27.20  & 282$^{2} $ & 107.91$^{b}$      &   780.49$^{g}$\\
TX Cam          & 05 00 50.39  & +56 10 52.60  & 179$^{2} $ & 91.15$^{b}$       &   778.79$^{g}$\\
U Cep       	& 01 02 18.45  & +81 52 32.08  & 390$^{3} $ & -{-}              &   583.91$^{g}$\\
U Her           & 16 25 47.47  & +18 53 32.86  & 461$^{5} $ & 30.16$^{a}$       &   255.54$^{i}$\\
UU For          & 02 37 23.07  & -26 58 42.31  & 710$^{3} $ & 36.85$^{a}$       &   -{-}        \\
UX Cyg          & 20 55 05.52  & +30 24 52.10  & 900$^{1} $ & 6.34$^{a}$        &   64.80$^{g}$ \\
VX Sgr          & 18 08 04.05  & -22 13 26.63  & 1570$^{4}$ & 626.32$^{a}$      &   3798.30$^{h}$\\
VY CMa      	& 07 22 58.33  & -25 46 03.24  & 502$^{2} $ & -{-}              &   6730.21$^{h}$ \\
W Hya           & 13 49 02.00  & -28 22 03.49  & 139$^{2} $ & 559.917$^{a}$     &   6146.00$^{h}$\\
W Per	        & 02 50 37.89  & +56 59 00.27  & 650$^{2} $ & 12.77$^{b}$       &   33.85$^{i}$\\
X Oph           & 18 38 21.12  & +08 50 02.75  & 235$^{2} $ & 65.78$^{a}$       &   -{-}        \\
Z Cas           & 23 44 31.59  & +56 34 52.70  & 797$^{3} $ & 3.98$^{b}$	    &   -{-}        \\
Z Cyg           & 20 01 27.50  & +50 02 32.69  & 930$^{3} $ & 3.37$^{a}$        &   -{-}        \\
\enddata
\tablenotetext{\alpha}{ Distance to Earth. Reference.- (1) \citet{Eng1979}; (2) \citet{Pic2010}; (3) \citet{Kim2014}; (4) \citet{Ric2012}; (5) \citet{Pal1993}}
\tablenotetext{\beta}{ Maser power denotes the integrated intensity of the maser emission. Reference for the SiO maser: (a) \citet{Kim2010}; (b) \citet{Cho2012};
 (c) \citet{Chi2016}; (d) \citet{Cho2009}; (e) Delingha, this work; (f) \citet{Nym1986}; (g) \cite{Par1998};
 (h) \citet{Wri1990}; (i) \citet{Ver012}}
\end{deluxetable}
\clearpage

\startlongtable
\Huge
\begin{deluxetable}{l|c|c|c|c}
\tablecaption{The silicate emission power of all the maser sources\label{tbl3}}
\tablehead{
\colhead{Name} & \colhead{$F_{\rm crys}$} & \colhead{$F_{\rm cont}$} & \colhead{$\frac{F_{\rm crys}}{F_{\rm am}+F_{\rm crys}}$} & \colhead{$F_{\rm 60\mum}$} \\
\colhead{}&\colhead{W/m$^{2}$ }&\colhead{W/m$^{2}$}& \colhead{}&\colhead{Jy}
}
\startdata
              AH Sco  &   2.195E-011   &  2.504E-010   &   0.081 &  73.310  \\
              BU And  &   3.486E-012   &  1.418E-011   &   0.197 &   7.100  \\
             Chi Cyg  &   2.011E-011   &  4.136E-010   &   0.046 &  80.670  \\
              EP Aqr  &   6.151E-012   &  8.459E-011   &   0.068 &  47.130  \\
              FP Aqr  &   7.518E-012   &  3.608E-011   &   0.172 &  17.320  \\
              GY Aql  &   4.735E-011   &  1.123E-010   &   0.296 &  47.310  \\
     IRAS 01037+1219  &   3.574E-011   &  2.281E-010   &   0.135 & 215.200  \\
     IRAS 05073+5248  &   1.064E-011   &  1.028E-010   &   0.094 &  72.390  \\
     IRAS 19192+0922  &   3.928E-012   &  2.479E-011   &   0.137 &  41.420  \\
               M ira  &   5.036E-011   &  6.327E-010   &   0.074 & 300.800  \\
               R And  &   2.491E-012   &  4.885E-011   &   0.049 &  24.160  \\
               R Aql  &   5.885E-012   &  9.727E-011   &   0.057 & 139.700  \\
               R Aqr  &   3.782E-011   &  2.166E-010   &   0.149 &  66.650  \\
               R Cas  &   9.526E-012   &  3.090E-010   &   0.030 & 102.800  \\
               R Hya  &   2.140E-011   &  2.130E-010   &   0.091 &  90.080  \\
               R Peg  &   1.457E-012   &  2.734E-011   &   0.051 &  11.060  \\
              RR Aql  &   8.530E-012   &  9.634E-011   &   0.081 &  27.470  \\
              RR Per  &   6.734E-013   &  1.103E-011   &   0.058 &   4.270  \\
              RT Vir  &   8.060E-012   &  7.520E-011   &   0.097 &  39.270  \\
               S Per  &   2.097E-011   &  1.048E-010   &   0.167 &  40.590  \\
               S Scl  &   8.244E-013   &  1.447E-011   &   0.054 &   6.930  \\
               S Vir  &   1.620E-012   &  1.721E-011   &   0.086 &   8.250  \\
              SS Peg  &   1.469E-012   &  9.323E-012   &   0.136 &   4.470  \\
               T Cas  &   3.413E-012   &  4.791E-011   &   0.067 &  24.380  \\
               T Cep  &   9.022E-012   &  1.181E-010   &   0.071 &  41.610  \\
              TX Cam  &   8.753E-012   &  1.523E-010   &   0.054 & 134.300  \\
               U Her  &   1.549E-012   &  3.589E-011   &   0.041 &  27.220  \\
              UU For  &   2.607E-012   &  3.755E-011   &   0.065 &  34.410  \\
              UX Cyg  &   9.143E-013   &  2.175E-011   &   0.040 &  43.940  \\
              VX Sgr  &   2.356E-011   &  1.011E-009   &   0.023 & 262.700  \\
              VY CMa  &   1.027E-010   &  3.588E-009   &   0.028 &1453.000  \\
               W Hya  &   3.502E-011   &  5.801E-010   &   0.057 & 195.000  \\
               W Per  &   2.035E-012   &  1.828E-011   &   0.100 &  14.870  \\
               X Oph  &   3.456E-012   &  5.258E-011   &   0.062 &  22.610  \\
               Z Cas  &   1.238E-012   &  1.524E-011   &   0.075 &   6.700  \\
               Z Cyg  &   2.520E-012   &  1.288E-011   &   0.164 &  10.660  \\
\enddata
\tablenotetext{}{ The total power of crystalline silicate dust(second column), continuum power(third column) and crystallinity of silicate dust(fourth column). }
\end{deluxetable}

\begin{figure}[!htp]
\centering
\includegraphics[height=12cm]{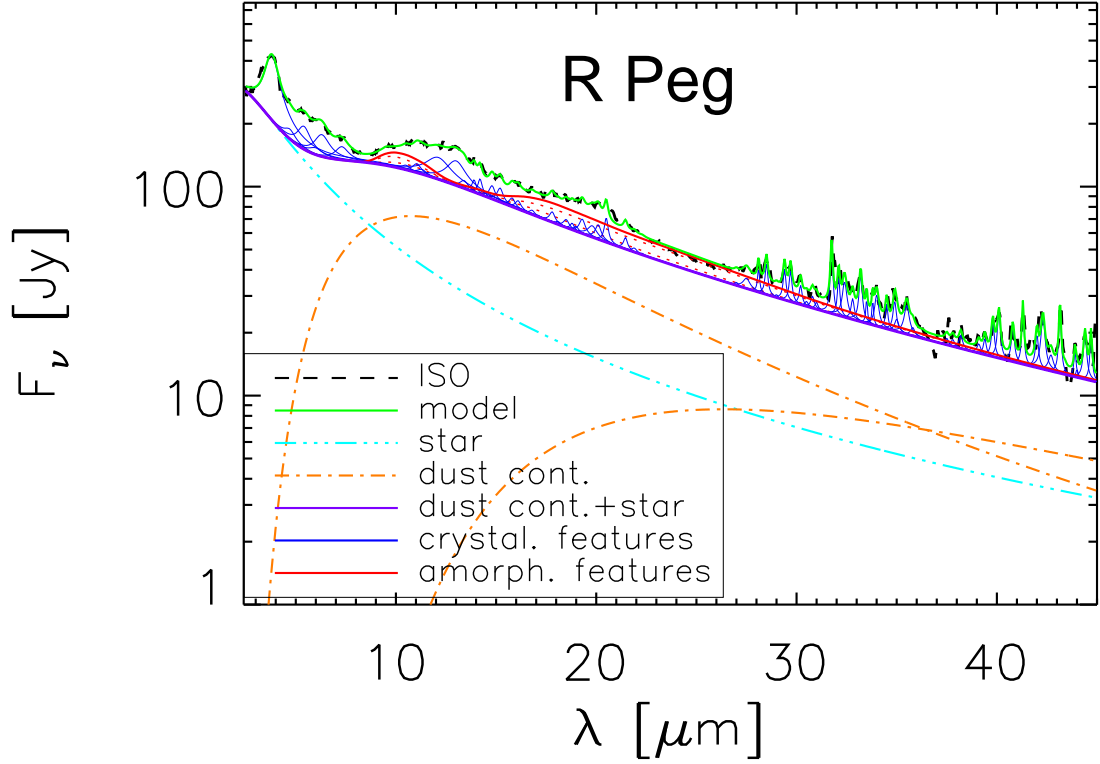}
\caption{The result of PAHFIT and the ISO spectra (dashed black line). This figure shows the dust continuum (orange lines), continuum emission of the central star (bright blue line), decomposed individual feature emissions (blue lines), the total continuum (dust continuum plus continuum emission of the central star, purple line) and the PAHFIT model result (green line).}
\label{Fig2}
\end{figure}

\begin{figure}[!htb]
\centering
   \includegraphics[width=12.0cm, angle=0]{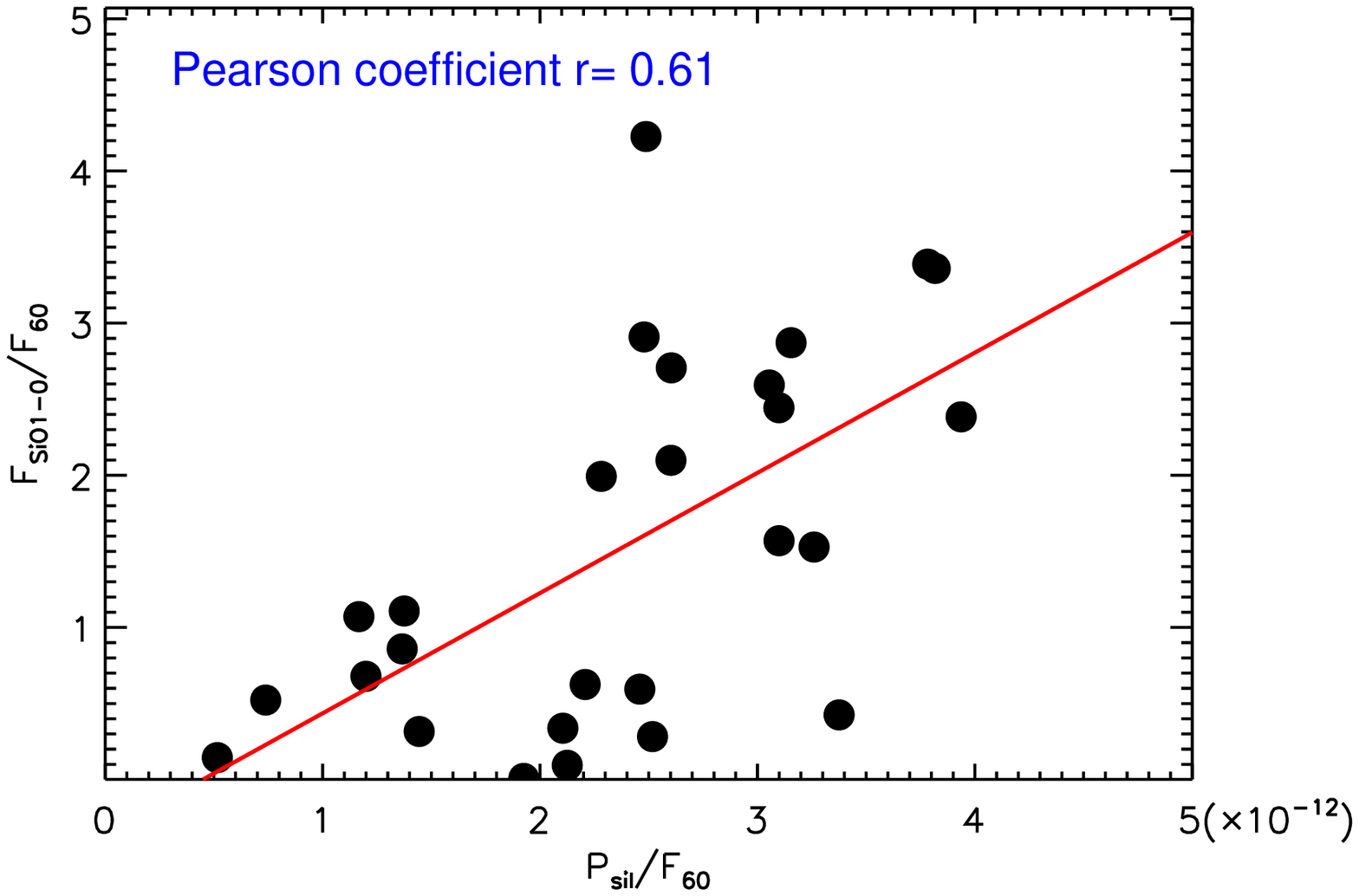}
   \includegraphics[width=12.0cm, angle=0]{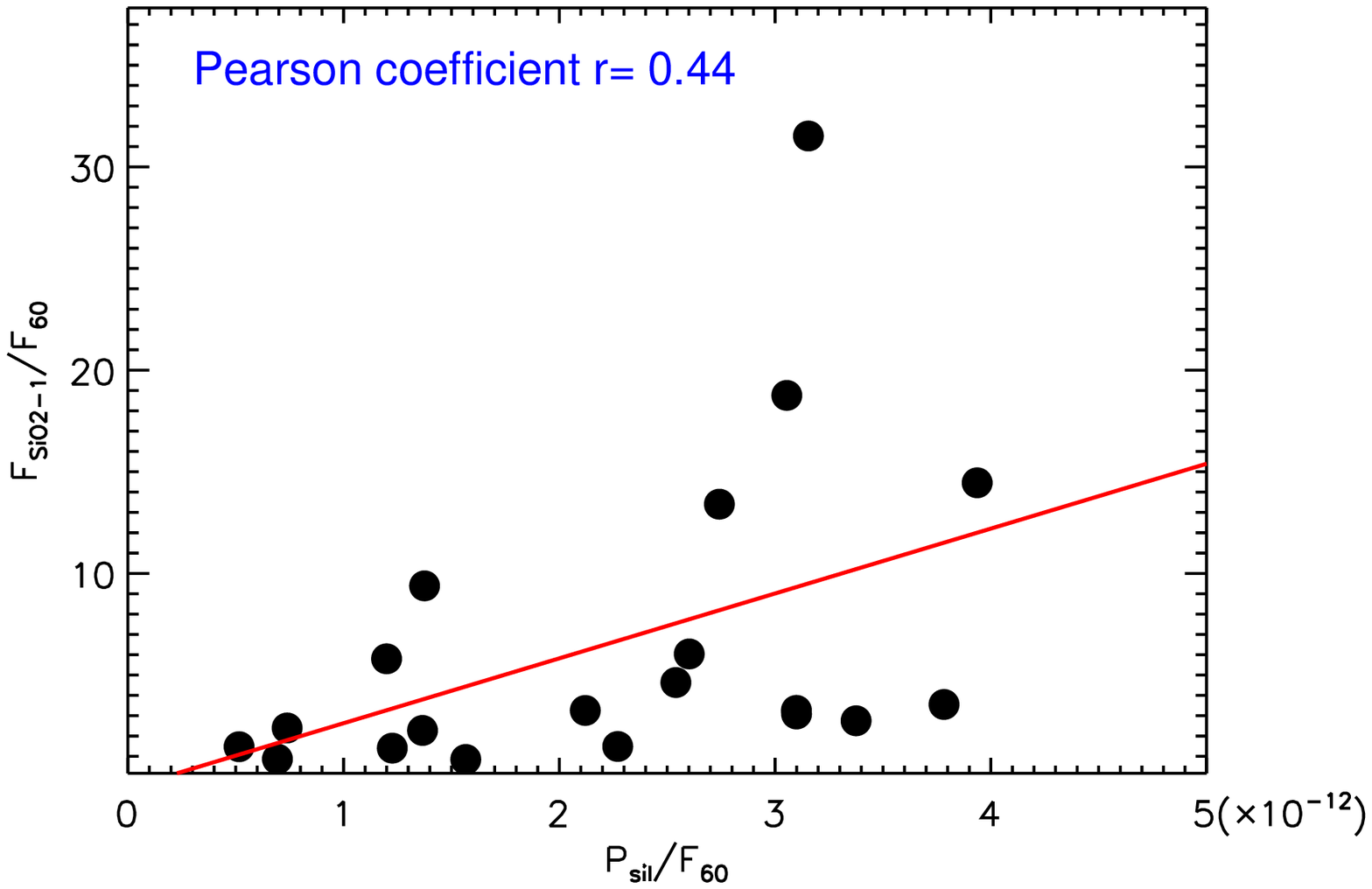}
   \caption{The relation of silicate emission power and SiO maser emission power. Upper panel is the relation of silicate to SiO $v=1,J=1-0$, while the lower panel shows the relation of silicate to SiO $v=1,J=2-1$. The red line is the linear fit of the data points and $r$ is the Pearson correlation coefficient.}
\label{Fig3}
\end{figure}

\begin{figure}
\centering
   \includegraphics[width=12.0cm, angle=0]{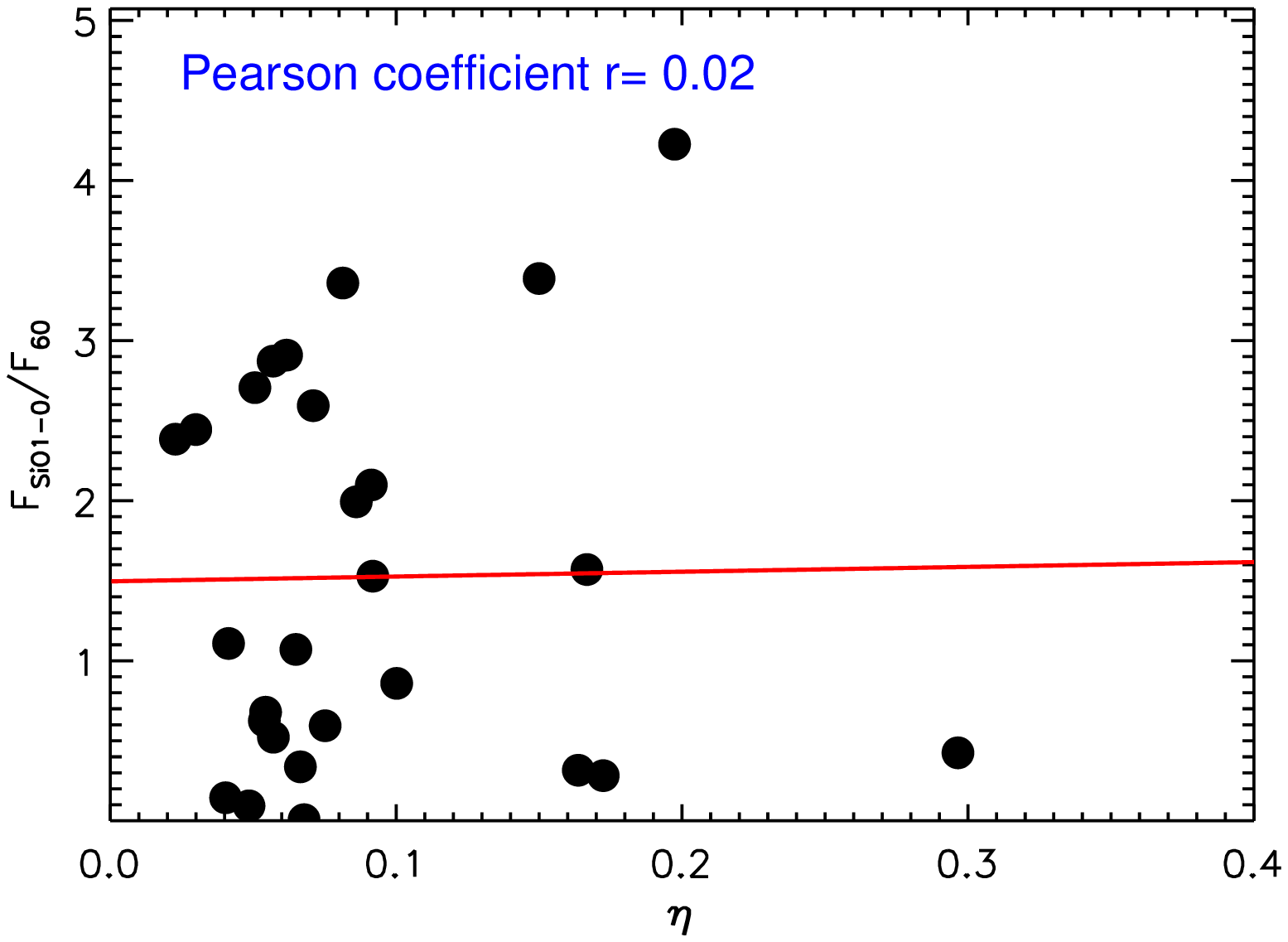}
   \includegraphics[width=12.0cm, angle=0]{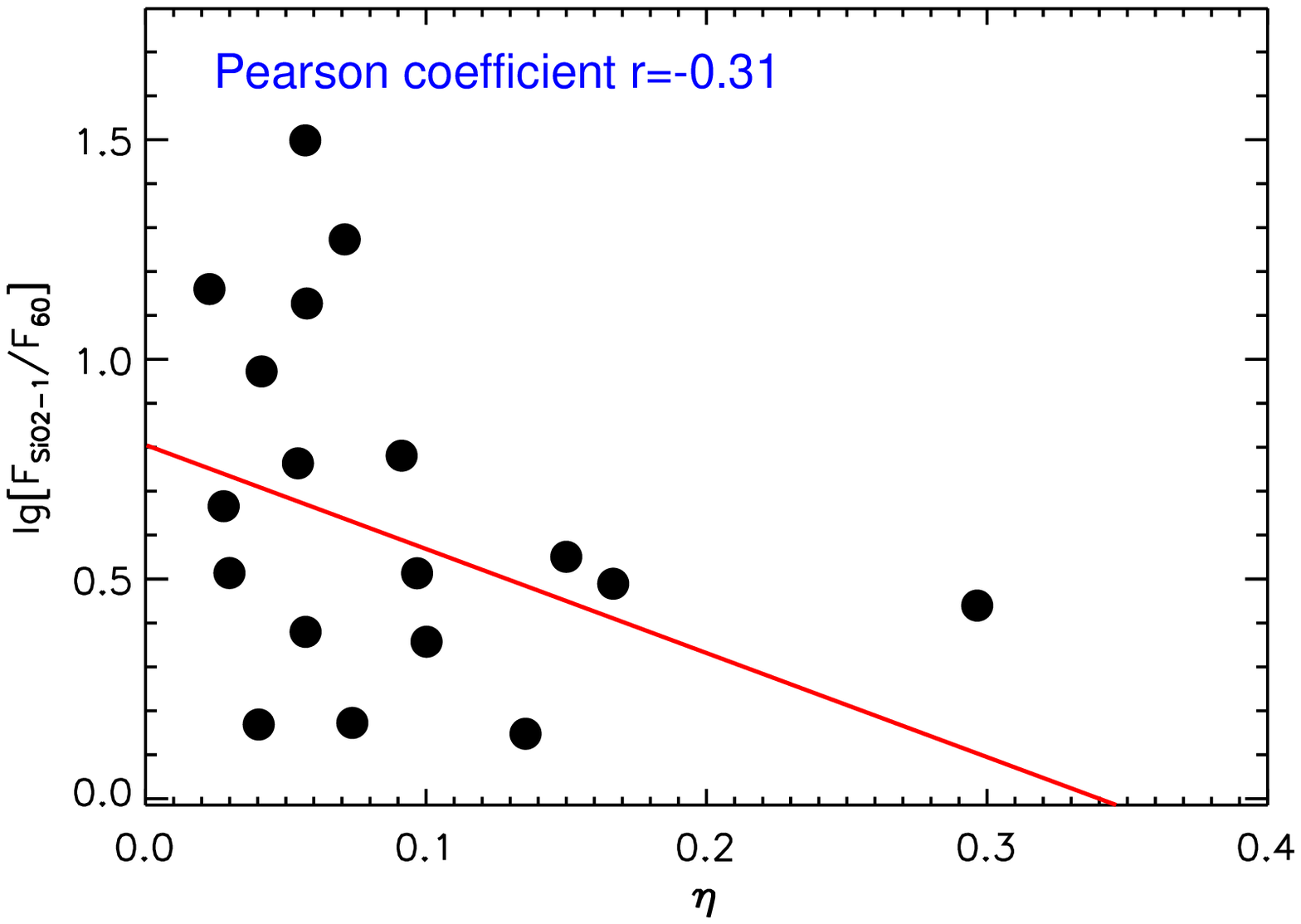}
   \caption{The relation of silicate crystallinity $\eta$ and SiO maser emission power. Upper panel is the relation of silicate to SiO $v=1,J=1-0$, while the lower panel shows the relation of silicate to SiO $v=1,J=2-1$. The red line is the linear fit of the data points and $r$ is the Pearson correlation coefficient.}
\label{Fig4}
\end{figure}



\end{document}